\documentclass[12pt]{article}
\usepackage{amssymb,amsfonts,amsmath}
\usepackage{latexsym}
\usepackage[dvips]{graphicx}

\newcommand{\spar}{\partial \hspace{-2.9mm}{\not}\hspace{2.9mm}}
\newcommand{\slaA}{A \hspace{-2.9mm}{\not}\hspace{2.9mm}}

\begin{document}
\title{ \large
PROPER-TIME FORMALISM
IN A CONSTANT MAGNETIC FIELD
AT FINITE TEMPERATURE AND CHEMICAL POTENTIAL
}
\author{TOMOHIRO INAGAKI
\and
{\it \normalsize
Information Media Center, Hiroshima University, 
Higashi-Hiroshima,}
\and
{\it \normalsize
Hiroshima, 739-8521, Japan} \\
{\it \normalsize
inagaki@hiroshima-u.ac.jp}
\and
DAIJI KIMURA$^*$ and TSUKASA MURATA$^\dagger$ 
\and
{\it \normalsize
Department of Physics, Hiroshima University, 
Higashi-Hiroshima,}
\and
{\it \normalsize
Hiroshima, 739-8526, Japan} \\
{\it \normalsize
$^*$kimura@theo.phys.sci.hiroshima-u.ac.jp} \\
{\it \normalsize
$^\dagger$murata@theo.phys.sci.hiroshima-u.ac.jp}}
\maketitle

\begin{abstract}
We investigate scalar and spinor field theories in a constant magnetic 
field at finite temperature and chemical potential.
In an external constant magnetic field the exact solution
of the two-point Green functions are obtained by using the 
Fock-Schwinger proper-time formalism. We extend it to the thermal 
field theory and find the expressions of the Green functions exactly 
for the temperature, the chemical potential and the magnetic field.
For practical calculations the contour of the proper-time integral 
is carefully selected. The physical contour is discussed in a constant 
magnetic field at finite temperature and chemical potential.
As an example, behavior of the vacuum self-energy is numerically 
evaluated for the free scalar and spinor fields. 
\end{abstract}

\newpage

\section{Introduction}
Some of the interesting cosmological and astrophysical 
situations are found in the state with high density, temperature 
and strong magnetic field. Neutron star is a dense object which 
has a large chemical potential. Recently it is observed 
that some of the neutron stars have extremely strong magnetic 
field.$^1$ A primordial magnetic field in the early universe 
is also interesting.$^{2,3}$
To understand the physics of such situation we consider the 
quantum field theory in an external magnetic field 
at finite chemical potential and temperature. 

One of the fundamental objects in the quantum field theory
is a two-point Green function. The exact form of the Green
function is necessary to deal with a strong magnetic field.
Much interest has been paid to obtain the Green function 
under external fields.
Schwinger found the exact expression of the Green function
in an external magnetic field by using the proper-time 
formalism in 1951.$^4$ 
The proper-time method is extended to deal with the thermal
system in Ref. 5. Trace of the Green function 
corresponds to the vacuum self-energy of the free field.
It is obtained by QED effective action and evaluated in a 
constant electromagnetic field at finite temperature$^{6,7}$
and at finite chemical potential.$^{8,9}$
Variety of approaches are used to discuss the contributions 
from both the temperature and the chemical potential in 
Refs. 10--13.

In the present paper the proper-time formalism is re-considered 
in the imaginary time form of the thermal field theory. We modify 
the formalism to introduce both the temperature and the chemical 
potential exactly. In most of previous analysis the proper-time 
integral was analytically performed by the Landau level expansion.
Since results are analytically continued to the wide range of 
parameters in the Landau level approach, these results may contain 
some approximation for the combined effect of temperature, chemical 
potential and external magnetic field. To obtain physical results 
we carefully choose the contour of the proper-time integral.$^{14}$ 
Here the explicit form of the scalar and fermion Green functions 
is written down in a proper-time form to discuss the physical contour 
in a constant magnetic field $H$ at finite temperature $T$ and 
chemical potential $\mu$. As an example, we numerically calculate 
the vacuum self-energy for the free scalar and fermion fields and 
discuss the contour dependence of the proper-time integral.

\section{Scalar Two-point Function at Finite $H$, $T$ and $\mu$}
First we study the Green function for a complex scalar
field in the constant magnetic field, $H$, at finite temperature, 
$T=1/\beta$, and chemical potential, $\mu$.
The chemical potential is defined for the global U(1) symmetry of the
complex field.$^{16}$
In the constant magnetic field, the Green function, $G(x,y;m_s)$, 
for the scalar field obeys the Klein-Gordon equation 
\begin{equation}
\left\{ (\partial_\mu + ieA_\mu)^2 + m^2_s \right\} G(x,y;m_s) 
    =\delta^4(x-y).
\label{Gs}
\end{equation}
We introduce the temperature and the chemical potential to this
equation. Following the standard procedure of the
imaginary time formalism,$^{15,16}$ the thermal Green function 
is defined in Euclidean space-time by
\begin{equation}
 \left\{ \left( i \partial_4 -i\mu + eA_4 \right)^2
       + \left( i \partial_j + eA_j \right)^2 + m^2_s \right\} G(x,y;m_s)
   = \delta^4_E( x-y ) ,
\label{Green}
\end{equation}
where $\delta^4_E(x-y) =\delta(x_4-y_4)
\delta^3(\boldsymbol{x}-\boldsymbol{y})$.
The time direction, $x_4$, is restricted between $0$ and $\beta$,
i.e. $x_4\in [0,\beta]$. 
Thus the thermal theory has no Lorentz invariance. It naturally
follows that the thermal equilibrium is defined along a specific time 
direction.

The Klein-Gordon equation is simplified by expanding the time 
direction, $x_4$, in Fourier series as
\begin{equation}
 G(x,y)=\frac1\beta 
    \sum_{n=-\infty}^\infty e^{-i \omega_n (x_4-y_4)}  
\widetilde{G}_n(\boldsymbol{x}, \boldsymbol{y}),
\label{G-expa}
\end{equation}
where the Matsubara frequency $\omega_n$ is given by $\omega_n=2\pi
n/\beta$ for a scalar field. 

Here we consider the constant magnetic field along the $z$-axis,
$A_\mu=\delta_{\mu 2}x_1H$, for simplicity. For the constant magnetic 
field the fourth component of $A_\mu$ vanishes, $A_4 =0$.
For a constant $A_\mu$ Eq.(\ref{Green}) reduces to
\begin{equation}
 \left\{ \left(\omega_n -i \mu \right)^2 
         - \left( \partial_j -i e A_j \right)^2 + m_s^2 \right\} 
              \widetilde{G}_n(\boldsymbol{x}, \boldsymbol{y}) = 
              \delta^3(\boldsymbol{x}-\boldsymbol{y}) .
\label{Green-Eu}
\end{equation}
The induced Green function, $\widetilde{G}_n$, has
similar form to the three dimensional Green function for the scalar
field with mass 
$M=\sqrt{(\omega_n -i \mu )^2 + m_s^2}$ in the electromagnetic field. 
$M$ develops an imaginary part only if both the temperature and the 
chemical potential have non-vanishing value.
Because of this imaginary part we must modify the original
Fock-Schwinger method as is shown below. 

We consider the proper-time Hamiltonian $H_n$;$^{4,17}$
\begin{equation}
 H_n = - \sum_{j=1}^3 (\partial_j -ie A_j)^2 
                    +( \omega_n - i\mu )^2 + m_s^2.
\label{Hami}
\end{equation} 
Evolution of the system is described by the proper-time, $\tau$. Then
the induced Green function $\widetilde{G}_n$ in Eq.(\ref{Green-Eu})
satisfies
\begin{equation}
 H_n \widetilde{G}_n(\boldsymbol{x}, \boldsymbol{y}) = 
 \delta^3(\boldsymbol{x}-\boldsymbol{y}).
\label{Green-pro}
\end{equation}
To solve this equation we introduce the unitary evolution operators 
$U^1_n$ and $U^2_n$ which are defined by
\begin{equation}
 i \frac\partial{\partial \tau}
 U^\alpha_n(\boldsymbol{x}, \boldsymbol{y}; \tau) 
 =H_n U^\alpha_n(\boldsymbol{x}, \boldsymbol{y}; \tau) , 
 \qquad({\rm \alpha=1,2}) ,
\label{Evol}
\end{equation}
with the boundary conditions
\begin{eqnarray}
 \lim_{\tau\to-\infty} U^1_n(\boldsymbol{x}, \boldsymbol{y}; \tau) 
    &=&  \lim_{\tau\to\infty} 
    U^2_n(\boldsymbol{x}, \boldsymbol{y}; \tau) =0 ,
\label{Bc1} \\
 \lim_{\tau\to-0} U^1_n(\boldsymbol{x}, \boldsymbol{y}; \tau) 
    &=&  \lim_{\tau\to+0} U^2_n(\boldsymbol{x}, \boldsymbol{y}; \tau)
 = \delta^3(\boldsymbol{x}-\boldsymbol{y}) .
\label{Bc2} 
\end{eqnarray} 
In the case of vanishing temperature or vanishing chemical potential
the induced Green function, $\widetilde{G}_n$, can be described by 
only one evolution operator, $U_n^1$. However, two types of the evolution 
operators with different boundary conditions are necessary to obtain a
finite $\widetilde{G}_n$ under non-vanishing temperature and chemical 
potential.
In the previous works contributions from the evolution operator 
$U^2_n$ is not completely considered.
The induced Green functions $\widetilde{G}_n(\boldsymbol{x},
\boldsymbol{y})$ are expressed by
$U^\alpha_n$ as
\begin{eqnarray}
 \widetilde{G}_n(\boldsymbol{x}, \boldsymbol{y}) = 
\begin{cases}
 \displaystyle
 -i \int^{-0}_{-\infty}\! d\tau ~U^1_n(\boldsymbol{x}, \boldsymbol{y}; \tau) 
 , \qquad {\rm (~n<0~),}  \\[2mm]
 \displaystyle
 i \int^\infty_{+0}\! d\tau ~U^2_n(\boldsymbol{x}, \boldsymbol{y}; \tau) 
, \qquad {\rm (~n>0~).}
\end{cases}
\label{GreenU}
\end{eqnarray}
After some straightforward calculations, see for example Refs. 17 
and 18, we obtain the evolution operators  
\begin{eqnarray}
 U^\alpha_n(\boldsymbol{x}, \boldsymbol{y}; \tau) 
  &=& \frac{ a^\alpha}{(4\pi)^{3/2} \left| \tau \right|^{3/2}}
        \frac{e H \tau}{\sin{(eH\tau)}} 
          \exp{\left\{ ie\int^{\boldsymbol{x}}_{\boldsymbol{y}} 
               d\xi\cdot A(\xi) \right\}}
\nonumber \\
  && \quad \times
    \exp{\left[\frac{i}4(x-y)_i eF_{ij} [\coth{(eF\tau)}]_{jk}
 (x-y)_k  \right. }
\nonumber \\ &&  \hspace{4cm} 
           -i\tau\left\{(\omega_n-i\mu)^2 + m_s^2 \right\}\Bigr],
\label{U1} 
\end{eqnarray}
where $F$ is the field strength and $H$ is the magnetic field.
$a^\alpha$ is defined by
\begin{eqnarray}
 a^\alpha =
\begin{cases}
 e^{+3\pi i/4}, \qquad {\rm (~\alpha=1~)},  \\
 e^{-3\pi i/4}, \qquad {\rm (~\alpha=2~)} .
\end{cases} 
\end{eqnarray}
The evolution operators, $U_n^\alpha$, are exponentially suppressed
at the limit $\tau \rightarrow - \infty$ for $n > 0$ and
$\tau \rightarrow \infty$ for $n < 0$.
It is clearly seen in Eq.(\ref{U1}) that the evolution operators
obey the boundary conditions (\ref{Bc1}) and (\ref{Bc2}).
Substituting Eqs.(\ref{GreenU}) and (\ref{U1}) in Eq.(\ref{G-expa}) 
we get the explicit expression of the Green function,
\begin{eqnarray}
G(x,y;m_s)  
 &=& \frac{-i}{\beta} \left[
       \sum_{n=-1}^{-\infty} e^{-i\omega_n (x_4-y_4)} 
         \int_{-\infty}^0\! d\tau ~U^1_n(\tau)\right.
\nonumber \\
 && \hspace{3cm} \left. 
         -\sum_{n=1}^{\infty} e^{-i\omega_n (x_4-y_4)} 
         \int^\infty_0\! d\tau ~U^2_n(\tau) \right] 
\nonumber \\
 &=& -\frac{i e^{3\pi i/4}}{(4\pi)^{3/2} \beta} 
   \sum_{n=1}^\infty e^{i\omega_n (x_4-y_4)}
\nonumber \\
 &&\quad \times
     \int_0^\infty\! d\tau
         \frac{eH}{\tau^{1/2}\sin{(eH\tau)}} 
           \exp{\left\{ ie 
           \int^{\boldsymbol{x}}_{\boldsymbol{y}}\! d\xi\cdot A(\xi) \right\}}
\nonumber \\
 &&\hspace{2cm}  \times \exp{\left[
     - \frac{i}4 (x-y)_i eF_{ij} [\coth{(eF\tau)}]_{jk} (x-y)_k 
      \right.}
\nonumber \\
 &&\hspace{3cm} 
         +i\tau\left\{ (\omega_n +i\mu)^2 + m_s^2 \right\}  \Bigr]  + (c.c.) .
\label{Green-s}
\end{eqnarray} 
The first term on the final result in Eq.(\ref{Green-s}) comes from the
integration of the evolution operator, $U^1_n$. The complex
conjugate of this, $(c.c.)$, appears from $U^2_n$ which is introduced 
to deal with the non-vanishing temperature and chemical potential. 

\section{Spinor Two-point Function at Finite $H$, $T$ and $\mu$}
Next we consider the Green function for a fermion field
in an external constant magnetic field  
at finite temperature and chemical potential.
The Green function,
$S(x,y;m_f)$, is defined by the Dirac equation;
\begin{eqnarray}
 \left( i\spar + e\slaA -i\mu\gamma_4 -m_f \right) S(x,y;m_f)
        = \delta^4_E(x-y),
\label{Green-fer}
\end{eqnarray}
where $m_f$ is the mass of the fermion field.

To calculate the analytical form of $S(x,y;m_f)$ it is more convenient
to introduce the bi-spinor function $G_f(x,y)$,
\begin{equation}
 S(x,y;m_f) =
 \left( i\spar + e\slaA -i\mu\gamma_4 +m_f \right) G_f(x,y).
\label{Green-G}
\end{equation}
The explicit form of $S(x,y;m_f)$ is determined by solving
the following equation for $G_f(x,y)$,
\begin{equation}
 \left\{ D^2 +\frac{i}2 eH(\gamma_1\gamma_2-\gamma_2\gamma_1)
        -2\mu\partial_4 +\mu^2 -m_f^2 \right\}G_f(x,y) =\delta^4_E(x-y),
\label{eq-Gre}
\end{equation}
where $D_j=\partial_j -ieA_j$. Here we choose the direction of the
constant magnetic field along the $z$-axis, $A_\mu =\delta_{\mu 2} x_1 H$. 
For a constant magnetic field the function,
$G_f(x,y)$, is expanded in Fourier series
\begin{equation}
 G_f(x, y)=\frac1\beta 
    \sum_{n=-\infty}^\infty e^{-i \omega_n (x_4-y_4)}  
 \widetilde{G}_n^f(\boldsymbol{x}, \boldsymbol{y}),
\label{Gf-expa}   
\end{equation}
and Eq.(\ref{eq-Gre}) reads
\begin{equation}
 \left\{ \sum_{j=1}^3  D_j^2
        +\frac{i}2 eH(\gamma_1\gamma_2-\gamma_2\gamma_1)
	  -(\omega_n - i\mu)^2 - m_f^2 \right\} 
 \widetilde{G}_n^f(\boldsymbol{x}, \boldsymbol{y}) 
 = \delta^3(\boldsymbol{x}-\boldsymbol{y}) .
\label{Gf-Eu}
\end{equation}
As in the scalar case, the induced function, $\widetilde{G}_n^f(x,y)$,
is calculated by introducing the proper-time Hamiltonian,
\begin{equation}
 H_n^f = \sum_{j=1}^3  D_j^2
        +\frac{i}2 eH(\gamma_1\gamma_2-\gamma_2\gamma_1)
	  -(\omega_n - i\mu)^2 - m_f^2.
\label{Hami-f}
\end{equation} 
According to the similar way in the scalar field we can find 
the bi-spinor function, $G_f(x,y)$. It is described by two types
of the unitary evolution operators $U^1_n$ and $U^2_n$ which satisfy
the boundary conditions (\ref{Bc1}) and (\ref{Bc2}).
Therefore the explicit expression of $G_f(x,y)$ is obtained by
\begin{eqnarray}
 G_f(x,y)
  &=& \frac{-i}{\beta} \left[
       \sum_{n=0}^{\infty} e^{-i\omega_n (x_4-y_4)} 
         \int_{-\infty}^{-0}\! d\tau ~U^1_n(\tau)\right.
\nonumber \\
 && \hspace{3cm} \left. 
         -\sum_{n=-1}^{-\infty} e^{-i\omega_n (x_4-y_4)} 
         \int^\infty_{+0}\! d\tau ~U^2_n(\tau) \right].
\label{solvG}
\end{eqnarray}
The Matsubara frequency for the fermion field is given by
$\omega_n=(2n+1)\pi/\beta$. The evolution operators 
$U^\alpha_n (\tau)$ are found to be
\begin{eqnarray}
 U^\alpha_n(\tau) 
 &=& \frac{b^\alpha}{(4\pi)^{3/2}|\tau|^{3/2}}\frac{eH\tau}{\sin (eH\tau)} 
  \exp
  \left\{ie\int^{\boldsymbol{x}}_{\boldsymbol{y}}\! d\xi\cdot A(\xi)\right\}
\nonumber \\
 && \times
    \exp{\left[ -\frac{i}4 (x-y)_i eF_{ij} [\coth{(eF\tau)}]_{jk}
    (x-y)_k  \right.}
\nonumber \\
 && \hspace*{30mm} \left. -i\tau\left\{ \frac12eF_{jk}\sigma_{jk} 
                    -(\omega_n -i \mu)^2 -m_f^2 \right\} \right],
\label{U2}
\end{eqnarray}
where $\displaystyle \sigma_{jk}=\frac{i}{2}[\gamma_j,\gamma_k]$, $F$ is 
the field strength and $b^\alpha$ is
\begin{eqnarray}
 b^\alpha =
\begin{cases}
 e^{-3\pi i/4}, \qquad {\rm (~\alpha=1~)},  \\
 e^{+3\pi i/4}, \qquad {\rm (~\alpha=2~)} .
\end{cases}
\end{eqnarray}
Inserting Eq.(\ref{U2}) into Eq.(\ref{solvG}), one easily derive the 
two-point Green function,
\begin{eqnarray}
&& S(x,y;m_f) = 
\nonumber \\
 && \hspace*{5mm}\left( i\spar + e\slaA -i\mu\gamma_4 +m_f \right) 
   \left( - \frac{i e^{-3\pi i/4}}{(4\pi)^{3/2}\beta}\right) 
    \sum_{n=0}^{\infty} e^{-i\omega_n (x_4-y_4)} 
\nonumber \\
 && \hspace*{5mm}\times
    \int^{\infty}_{0}\! d\tau \frac{eH}{\tau^{1/2}\sin (eH\tau)}
    \exp\left\{ 
    ie\int^{\boldsymbol{x}}_{\boldsymbol{y}}\! d\xi\cdot A(\xi) 
    \right\}
\nonumber \\
 && \hspace*{10mm}\times
    \exp{\left[ \frac{i}4 (x-y)_i eF_{ij} [\coth{(eF\tau)}]_{jk}
    (x-y)_k  \right.}
\nonumber \\
 && \hspace*{15mm}
  \left. 
     +i\tau\left\{ \frac12eF_{jk}\sigma_{jk} 
                    -(\omega_n -i \mu)^2 -m_f^2 \right\} \right]
 + (c.c.) . 
\label{Green-f}
\end{eqnarray}
The complex conjugate term, (c.c.), in Eq.(\ref{Green-f}) comes 
from $U^2_n$ which is introduced to deal with effects of both
the temperature and the chemical potential. 

\section{Behavior of the Vacuum Self-energy}
At the one loop level the vacuum self energy for a free field is 
given by the trace of the Green function. Here we calculate it
for a free scalar and a free fermion fields at finite $H$, $T$ and 
$\mu$. 

For a free scalar field with mass $m_s$ the trace of the Green 
function (\ref{Green-s}) 
becomes
\begin{eqnarray}
 \frac1{\beta V} {\rm Tr}G (x, x)
  &=& \frac{e^{\pi i/4}}{(4\pi)^{3/2} \beta} \sum_{n=1}^\infty
      \int_0^\infty\! d\tau \frac{eH}{\tau^{1/2} \sin{(eH\tau)}}
\nonumber \\
  && \hspace{1.4cm} \times
        \exp{\left[ i\tau\left\{(\omega_n+i\mu)^2 + m_s^2 \right\} \right]}
    + (c.c.) ,
\label{Tr-Green1}
\end{eqnarray}
where $V$ is the 3-dimensional volume.
If it is larger than $m^2_s$, naive perturbation loses validity to 
evaluate the radiative correction in a scalar theory with interactions.
To get rid of this difficulty we must use a resumed propagator known 
as the ring diagram resummation in the thermal field theory.$^{16}$

Carrying out the trace of the Green function (\ref{Green-f}) 
with respect to space-time and spinor legs, we find the vacuum 
self-energy for a free fermion field,
\begin{eqnarray}
 \frac1{\beta V} {\rm Tr}S (x, x)
  &=& \frac{4e^{3\pi i/4} m_f}{(4\pi)^{3/2} \beta} \sum_{n=0}^{\infty}
      \int_0^\infty\! d\tau \frac{eH \cot{(eH\tau)}}{\tau^{1/2} }
\nonumber \\
  && \hspace{1.4cm} \times
        \exp{\left[ -i\tau
        \left\{(\omega_n -i\mu)^2 +m_f^2 \right\} \right]}
    + (c.c.) .
\label{Tr-Green2}
\end{eqnarray}
For the vanishing temperature and/or chemical potential
the contribution from the evolution operator $U^2_n$ is coincide
with the one from $U^1_{-n}$ and contour dependence of the 
proper-time integral disappears.
Therefore our result must be coincide with the one found in the previous 
work. For $\mu \rightarrow 0$ Eq.(\ref{Tr-Green2}) exactly agrees 
with the one obtained in Ref. 6. At the limit, 
$T \rightarrow 0$ $(\beta \rightarrow \infty)$, Eq.(\ref{Tr-Green2}) 
reproduce the results obtained in Refs. 8 and 9. 

Performing the proper-time integral and the summation 
in Eq.(\ref{Tr-Green1}) and Eq.(\ref{Tr-Green2}) numerically,
we evaluate the vacuum self-energy for the free scalar and the 
free fermion fields. We choose the contour of proper-time 
integration in the first term of the right hand side of 
Eq.(\ref{Tr-Green1}) slightly above the real axis.
As is known, this contour gives physical results at the limit 
$\mu \rightarrow 0$ and $T \rightarrow 0$.$^4$ 
It is natural to take the same contour at finite $\mu$ and $T$.

\begin{figure}[ht]
  \begin{center}
   \begin{tabular}{cc}
    \resizebox{60mm}{!}{\includegraphics{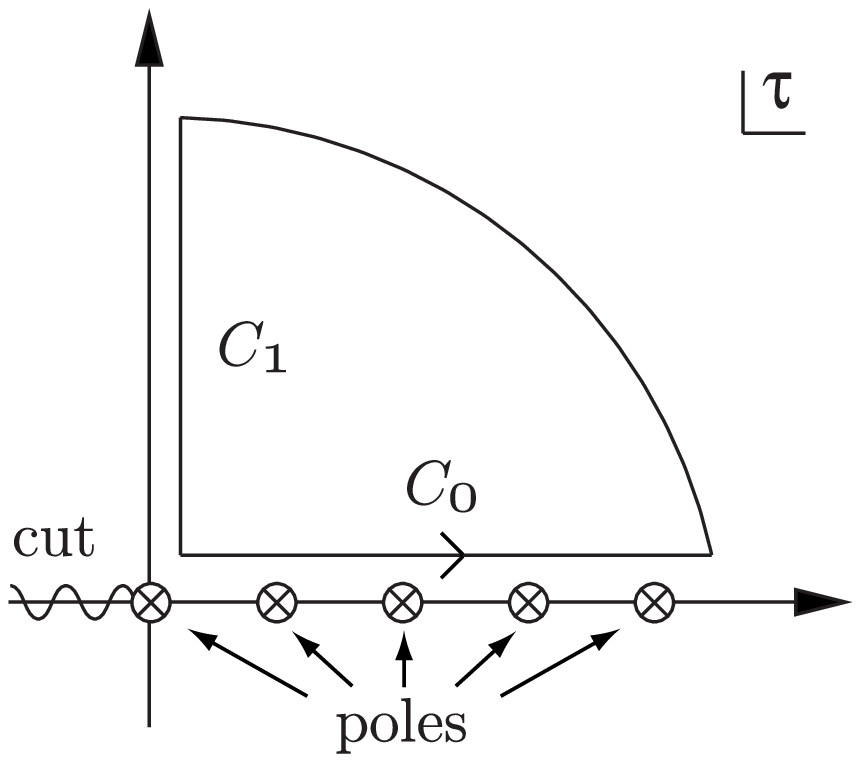}}
    &
    \resizebox{60mm}{!}{\includegraphics{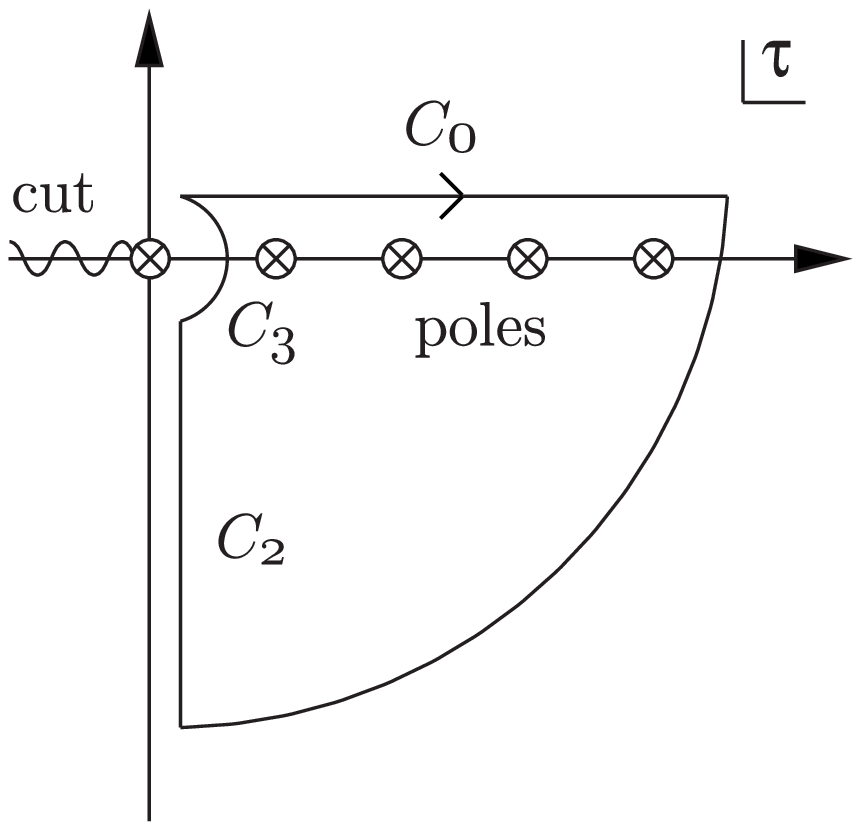}}
    \\
    {(a)} & {(b)}
   \end{tabular}
  \label{path}
  \caption{Contour on the complex $\tau$ plane.}
  \end{center}
\end{figure}

For $\omega_1^2-\mu^2+m_s^2 > 0$ the integrand is exponentially suppressed 
at the infinity above the real axis. We close the contour as is 
shown in Fig. \ref{path} (a) and perform the proper-time integral 
along the path $C_1$. The complex conjugate of the result gives
the second term in the right hand side in Eq.(\ref{Tr-Green1}).
Thus the ${\rm Tr}G(x,x)$ is found to be
\begin{eqnarray}
 \frac{1}{\beta V} {\rm Tr}G(x,x) 
  =\frac{1}{4\pi^{3/2}\beta} \sum_{n=1}^\infty \int_{1/\Lambda^2}^\infty
  d\tau f_s(\tau, n),
\end{eqnarray}
where $f_s(\tau, n)$ is 
\begin{equation}
 f_s(\tau, n)=\frac{eH}{\sqrt{\tau}} \frac{\cos(2\omega_n\mu\tau)}
  {\sinh(eH\tau)} e^{-\tau(\omega_n^2-\mu^2+m_s^2)} . 
\label{C3}
\end{equation}
Here we introduce the proper-time cut-off $\Lambda$ to regularize
the theory. 

In the case $\omega_1^2-\mu^2+m_s^2<0$ we must consider two kinds of paths
in Fig. \ref{path} (a) and (b). For a positive $\omega_n^2-\mu^2+m_s^2$ 
the integrand drops at the infinity above the real axis. 
If $\omega_n^2-\mu^2+m_s^2$ is negative, the integrand is suppressed at 
the infinity below the real axis. We calculate the proper-time integral
along the paths $C_1$, $C_2$, $C_3$ and add the contribution from the poles 
on the real axis. After some calculations ${\rm Tr}G(x,x)$ is obtained by
\begin{eqnarray}
 \frac{1}{\beta V} {\rm Tr}G(x,x) 
  &=&\frac{1}{4\pi^{3/2}\beta} \sum_{n>[N]}^\infty 
  \int_{1/\Lambda^2}^\infty d\tau f_s(\tau, n) \nonumber \\ 
  &&+\frac{1}{4\pi^{3/2}\beta} \sum_{n=1}^{[N]} \left[
   h_{s0}(n)+h_{sj}(n)-\int_{1/\Lambda^2}^\infty d\tau g_s(\tau, n)
   \right] ,
 \end{eqnarray}
where $N=\beta\sqrt{\mu^2-m_s^2}/(2\pi)$, $[N]$ is the Gauss notation 
and
\begin{eqnarray}
 g_s(\tau, n)&=&\frac{eH}{\sqrt{\tau}} \frac{\sin(2\omega_n\mu\tau)}
  {\sinh(eH\tau)} e^{-\tau(\mu^2-\omega_n^2-m_s^2)}, \label{D3} \\
 h_{s0}(n)&=&\frac{e^{-\pi i/4}}{2} \int_{-\pi/2}^{\pi/2} d\theta 
  \frac{eH}{\Lambda} \frac{e^{i\theta/2}}{\sin(eHe^{i\theta}/\Lambda^2)}
  \nonumber \\
  &&\times\exp[i\{(\omega_n+i\mu)^2+m_s^2\}e^{i\theta}/\Lambda^2]+(c.c.), 
  \label{D4}\\
 h_{sj}(n)&=&2\sqrt{\pi} \sum_{l=1}^{\infty} (-1)^l 
  \left(\frac{eH}{l}\right)^{1/2} e^{-2\pi l\omega_n\mu/(eH)} \nonumber \\ 
  &&\times \cos\left[\frac{\pi l}{eH} (\mu^2-\omega_n^2-m_s^2)
  +\frac{\pi}{4} \right]. \label{Res} 
\end{eqnarray}
$h_{s0}(n)$ is the contribution from the path $C_3$ around the pole
at $\tau=0$ and $h_{sj}(n)$ is sum of residues at $eH \tau=\pi l$.

In the case of the fermion field the above contour gives the physical
result at $\mu \rightarrow 0$ and $T \rightarrow 0$ for the second
term ``$(c.c.)$'' of right hand side of Eq.(\ref{Tr-Green2}).
According to the similar analysis with the scalar field, 
we calculate the ${\rm Tr}S(x,x)$. For a positive 
$\omega_0^2-\mu^2+m_s^2$ the proper-time integral is performed
along the path $C_1$,
\begin{eqnarray}
 \frac{1}{\beta V} {\rm Tr}S(x,x) 
  =-\frac{m_f}{\pi^{3/2}\beta} \sum_{n=0}^\infty \int_{1/\Lambda^2}^\infty
  d\tau f(\tau, n),
\end{eqnarray}
where $f(\tau, n)$ is
\begin{equation}
  f(\tau, n)=\frac{eH}{\sqrt{\tau}} \coth(eH\tau) \cos(2\omega_n\mu\tau)
  e^{-\tau(\omega_n^2-\mu^2+m_f^2)} .
\end{equation}
For a negative $\omega_0^2-\mu^2+m_s^2$ we evaluate the proper-time 
integral along the path $C_1$, $C_2$, $C_3$ and consider the influence
from pole on the real axis.
\begin{eqnarray}
 \frac{1}{\beta V} {\rm Tr}S(x,x) 
  &=&-\frac{m_f}{\pi^{3/2}\beta} \sum_{n>[N]}^\infty 
  \int_{1/\Lambda^2}^\infty d\tau f(\tau, n) \nonumber \\ 
  &&+\frac{m_f}{\pi^{3/2}\beta} \sum_{n=0}^{[N]} \left[
   h_0(n)+h_j(n)+ \int_{1/\Lambda^2}^\infty d\tau g(\tau, n)
  \right],
\end{eqnarray}
where $N=\beta\sqrt{\mu^2-m_f^2}/(2\pi)-1/2$, $g(\tau, n)$, $h_0(n)$ 
and $h_j(n)$ are given by
\begin{eqnarray}
 g(\tau, n)&=&\frac{eH}{\sqrt{\tau}} \coth(eH\tau) \sin(2\omega_n\mu\tau)
  e^{-\tau(\mu^2-\omega_n^2-m_f^2)}, \\
 h_0(n)&=&\frac{e^{3\pi i/4}}{2} \int_{-\pi/2}^{\pi/2} d\theta 
  \frac{eH}{\Lambda} e^{i\theta/2}\cot(eHe^{i\theta}/\Lambda^2)
  \nonumber \\
  &&\times\exp[i\{(\omega_n+i\mu)^2+m_f^2\}e^{i\theta}/\Lambda^2]+(c.c.) ,\\
 h_j(n)&=&2\sqrt{\pi} \sum_{l=1}^{\infty}  
  \left(\frac{eH}{l}\right)^{1/2} e^{-2\pi l\omega_n\mu/(eH)} \nonumber \\ 
  &&\times \sin\left[\frac{\pi l}{eH} (\mu^2-\omega_n^2-m_f^2)
  -\frac{\pi}{4} \right]. 
\end{eqnarray}
$h_{0}$ corresponds to the contribution from the path $C_3$ 
and $h_{j}$ is sum of residues at $eH\tau=\pi l$.

\begin{figure}
  \begin{center}
   \begin{tabular}{cc}
    \resizebox{!}{62mm}{\includegraphics{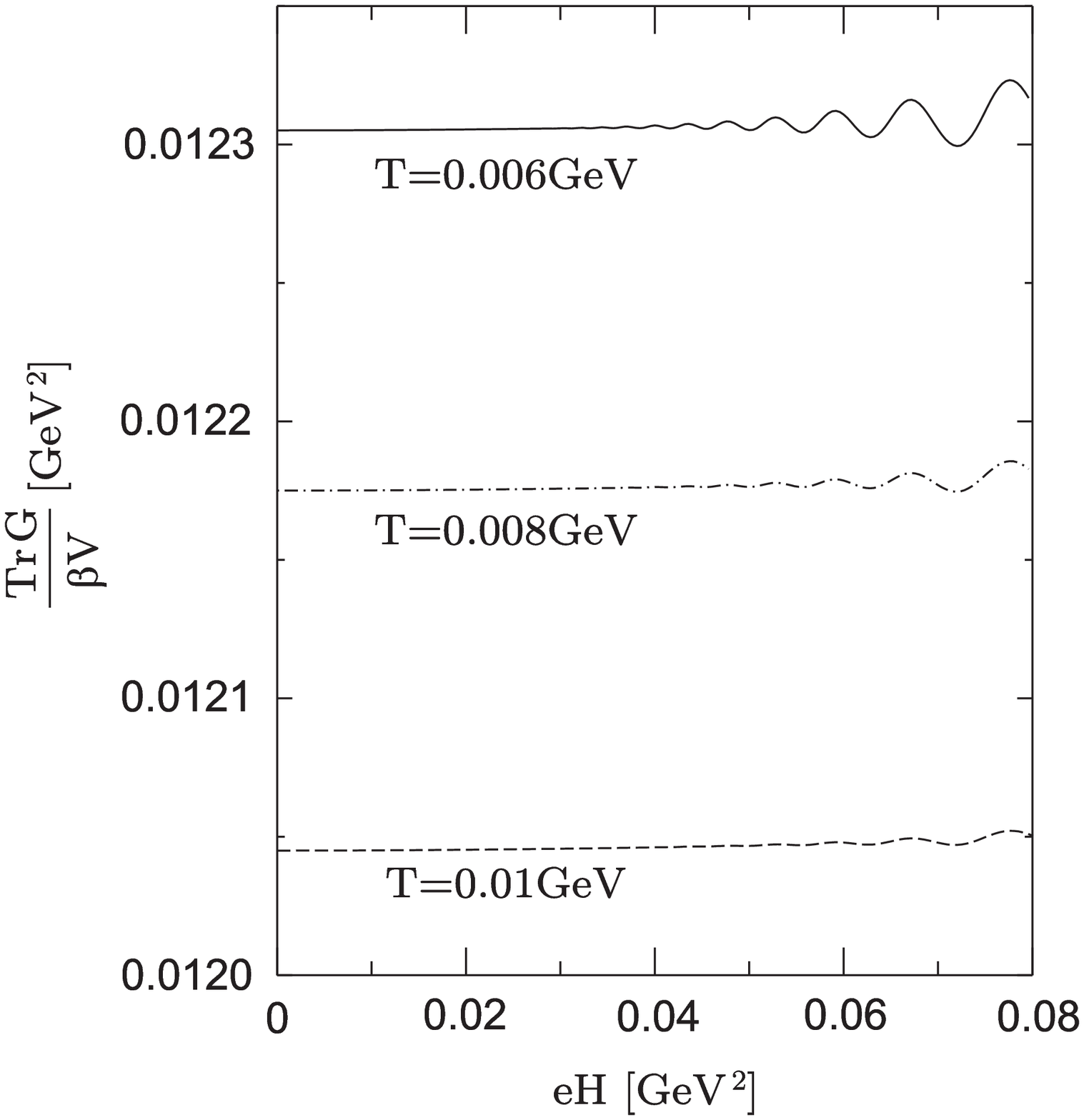}} 
    &
    \resizebox{!}{62mm}{\includegraphics{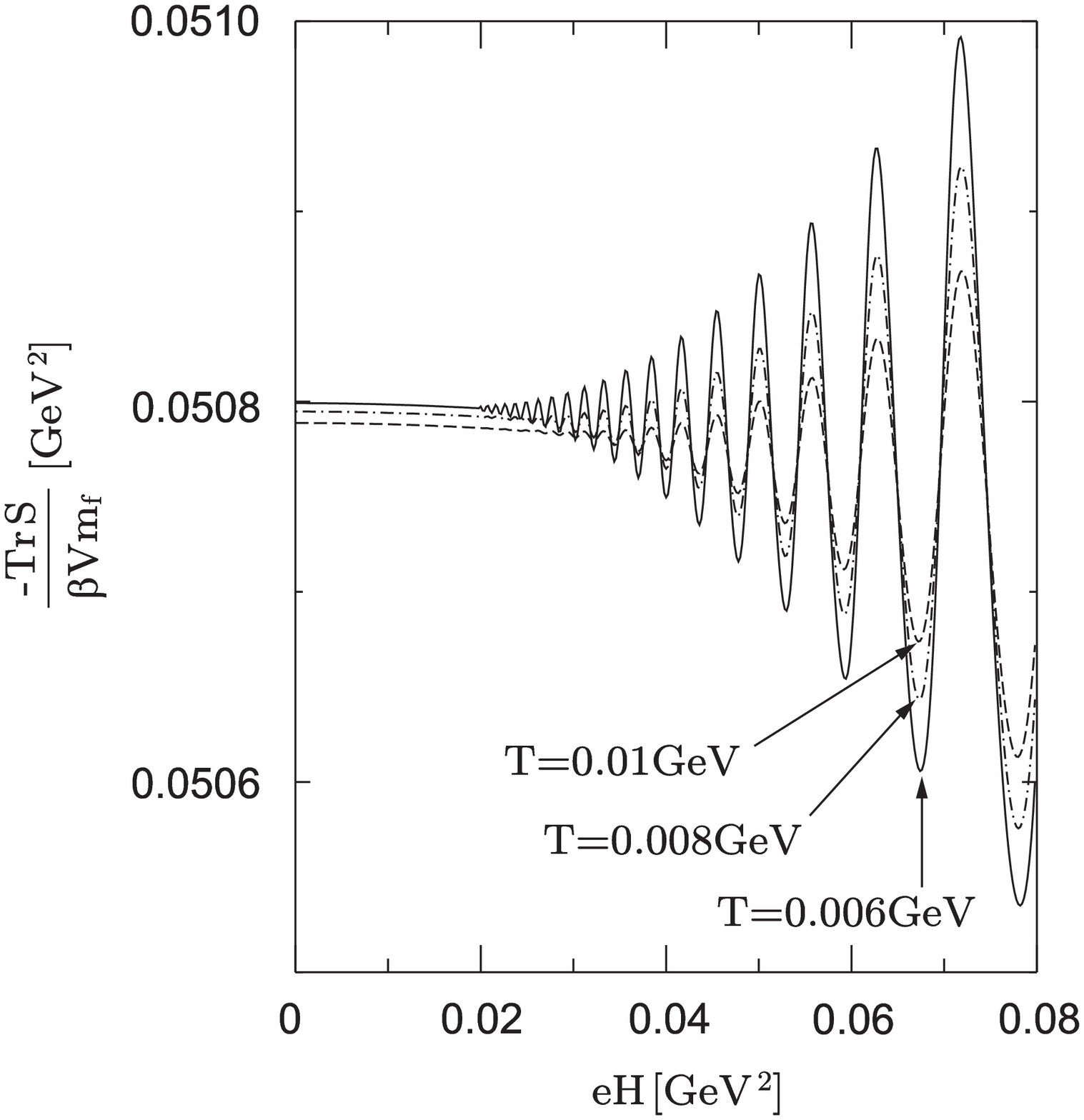}}
    \\
    scalar & fermion
   \end{tabular}
 \caption{Behaviors of the vacuum self energy as a function of 
  $H$ with $T$ and $\mu$ fixed. We set $\Lambda=2$GeV,
  $m_s=m_f=0.1$GeV, $\mu=1$GeV and
  $T=0.006$GeV, $T=0.008$GeV, $T=0.01$GeV.}
 \label{eB}
\vglue 4ex
   \begin{tabular}{cc}
    \resizebox{!}{62mm}{\includegraphics{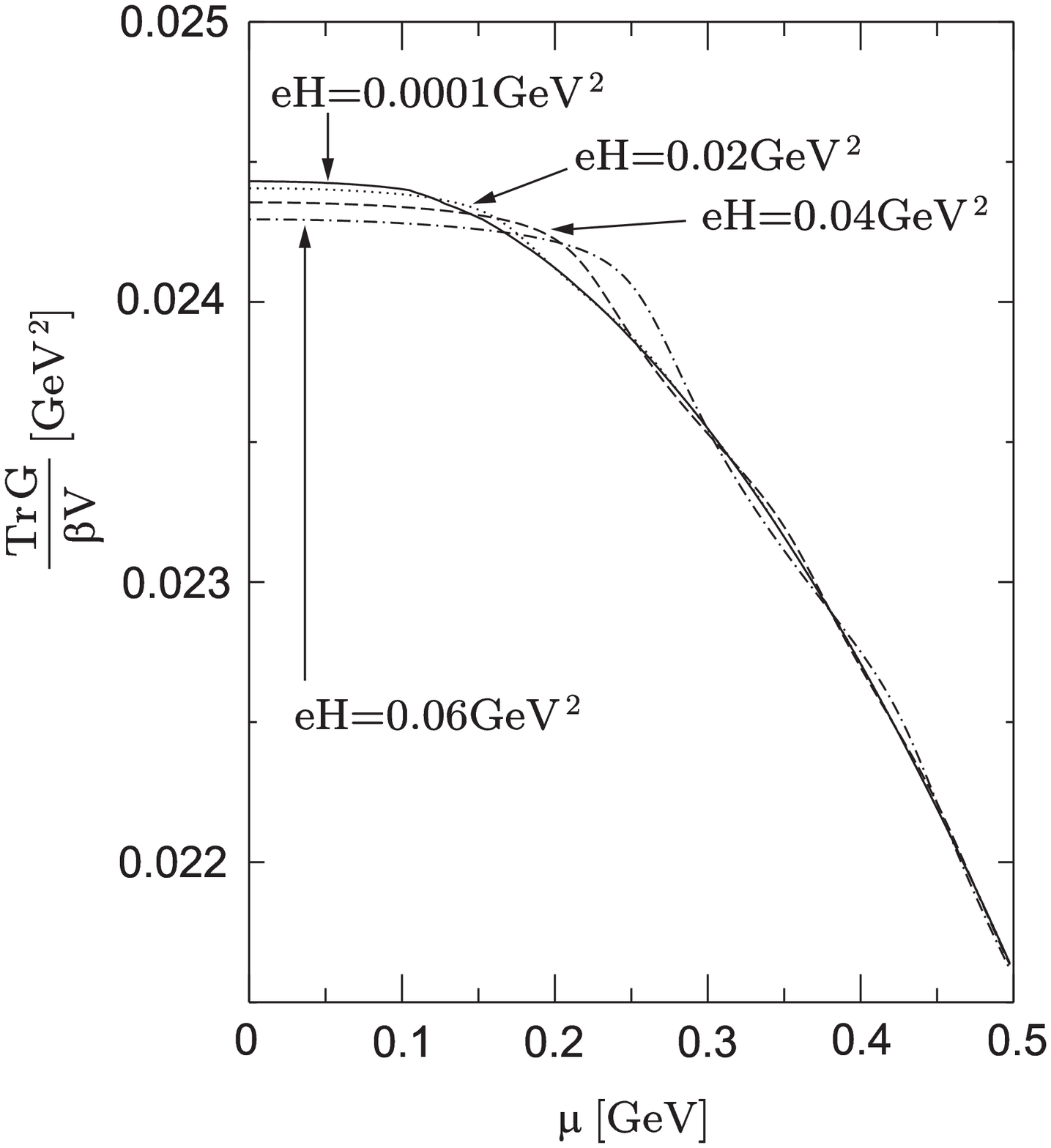}}
    &
    \resizebox{!}{62mm}{\includegraphics{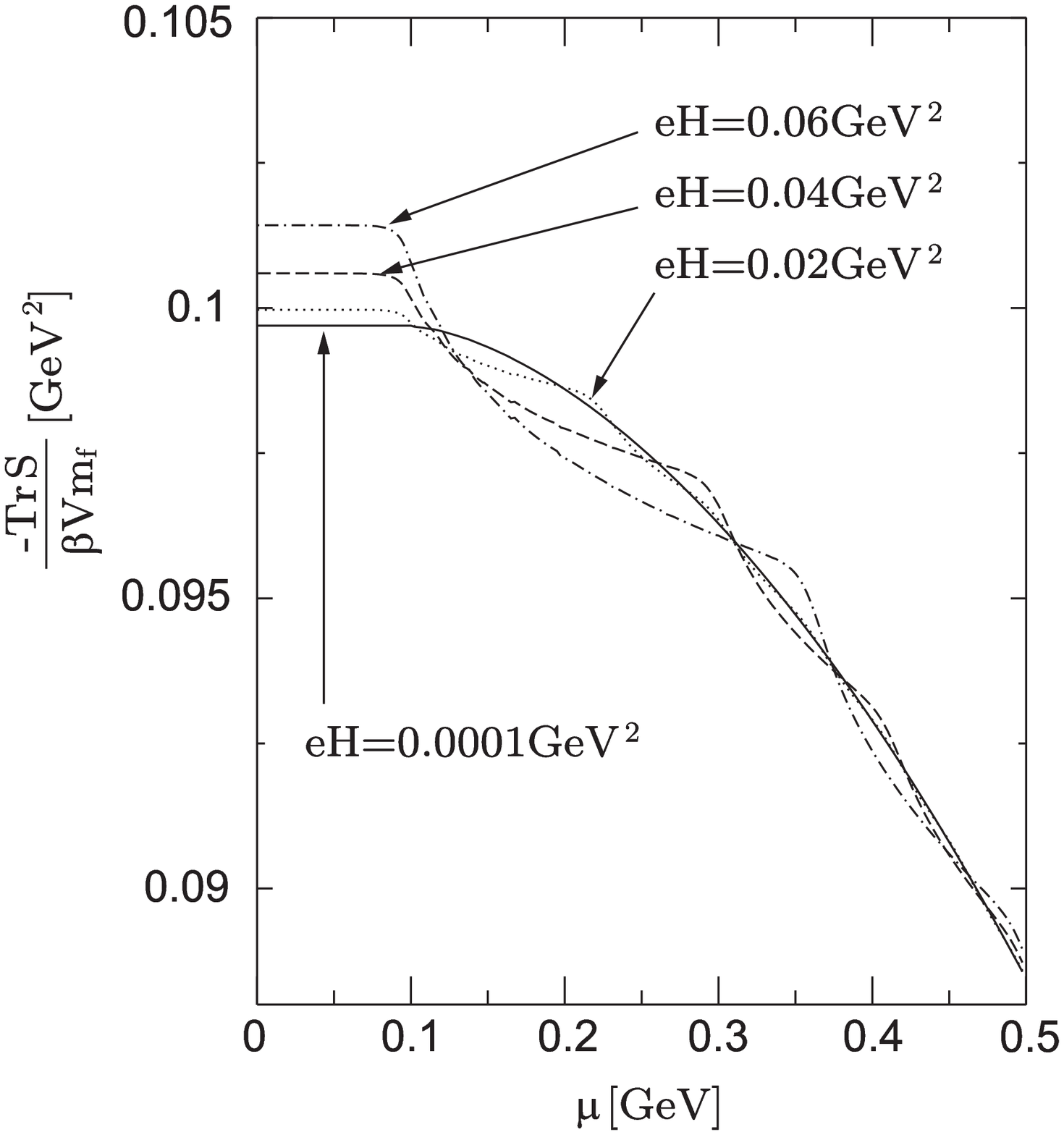}}
    \\
    scalar & fermion
  \end{tabular}
 \caption{Behaviors of the vacuum self energy as a function of 
  $\mu$ with $H$ and $T$ fixed. We set $\Lambda=2$GeV,
 $m_s=m_f=0.1$GeV, $T=0.006$GeV and
  $eH=0.0001$GeV$^2$, $eH=0.02$GeV$^2$, $eH=0.04$GeV$^2$, 
  $eH=0.06$GeV$^2$.}
 \label{mu}
 \end{center}
\end{figure}

Performing the integration over $\tau$ and the summation numerically,
we obtain the behaviors of the trace of the Green functions, i.e. 
the vacuum self energy at the one loop level.
In Fig. \ref{eB} we show behaviors of the vacuum self-energy 
as a function of the external magnetic field $H$ with $T$ and $\mu$ 
fixed. 
For a case of neutron star many interest has been payed to the QCD 
phase structure. Thus we suppose 
the proper-time cut-off is more than QCD scale,
$\Lambda_{\rm QCD} < \Lambda \sim 2$GeV. The scalar and fermion mass
is taken to be the pion mass scale, $m_s \sim m_f \sim 0.1$GeV.
The present upper limit for the magnetic field in the neutron star is
of the order, $eH \sim O(0.01{\rm GeV}^2)$.$^{19}$
An oscillating mode is observed for both the scalar and 
fermion field. The amplitude of the oscillation becomes larger
as $H$ increases and/or $T$ decreases. The oscillation disappears 
for higher temperature.
For the neutron star the upper limit of the magnetic
field, $eH$, is of the order $O(0.01\mbox{GeV}^{-2})$.
It seems to be difficult to see the magnetic oscillation
appeared in Fig. \ref{mu} unless the neutron star is extremely
cold. It agrees with the result obtained in Ref. 11.

Behaviors of the vacuum self-energy is illustrated as a function of 
$\mu$ with $T$ and $H$ fixed in Fig. \ref{mu}. The trace of the 
two-point function goes down as $\mu$ increases for both the scalar 
and fermion field. We can see the oscillating mode for both cases.
For the fermion field the mode is the origin of the van Alphen-de 
Haas magnetic oscillations as is shown in Ref. 11.
Such a effect is not found in the scalar field theory, since
the scalar field has no sharp Fermi surface.

As in known, the trace of the scalar two-point function contains a 
term which is proportional to $T^2$.$^{16}$ 
Indeed, Eq.(\ref{C3}) reduces to the well-known result,
\begin{equation}
\frac{{\rm Tr}G}{\beta V} = 2 \frac{T^2}{24}+\mbox{O}(T) ,
\label{ring}
\end{equation}
at the limit $\mu, H$ and  $m_s\rightarrow 0$.
To obtain it we drop the surface term of the proper-time integral 
and use a formula $\zeta(-1)=-1/12$.
$T^2$ behavior is not observed in Fig. \ref{eB}, because the
surface term is proportional to $\Lambda T$. The temperature 
considered here is too small compared with the cut-off scale
$\Lambda$. $T$-dependence of the surface term is canceled out
if we take the T-dependent cut-off, $\Lambda \propto 1/T$. 
Similar property is found in the momentum cut-off 
regularization for $H=0$.


\section{Conclusion}
We have investigated the scalar and fermion field theories
at finite temperature, chemical potential and constant magnetic 
field. The explicit expressions of the two-point Green functions
are found by using the proper-time formalism. If both the 
temperature and the chemical potential exist, we must modify
the Fock-Schwinger proper-time method by introducing two types 
of the evolution operators with different boundary conditions.

The proper-time integrations remain in our final expressions of
the two-point Green functions $G(x,y;m_s)$ and $S(x,y;m_f)$.
The remained integrand is exponentially suppressed
at the limit $\tau \rightarrow \infty$. There are poles at
$eH\tau = n\pi$ for any integer $n$. 
Because of these poles the naive Wick rotation has no validity.
\footnote{After the naive Wick rotation $\tau \rightarrow it$, 
we can perform the summations in Eq.(\ref{Tr-Green1}) and 
Eq.(\ref{Tr-Green2}) and the vacuum self-energy
is described by the elliptic theta function,$^{13}$
However, it is not valid in our formalism.}
We carefully take the contour of proper-time integration slightly 
above the real axis to avoid these poles in the complex $\tau$ 
plane. In the case with large chemical potential we must consider
two kinds of contour. One of the contour contains the poles at
$eH\tau = n\pi$. On the other hand, the naive analytic 
continuation from the low chemical potential takes the contour 
below the real axis for $\omega_n - \mu^2 + m^2 < 0$. Thus it 
drops the contribution from the poles on the real axis.
It gives only approximate results.

\begin{figure}
  \begin{center}
    \resizebox{!}{6.2cm}{\includegraphics{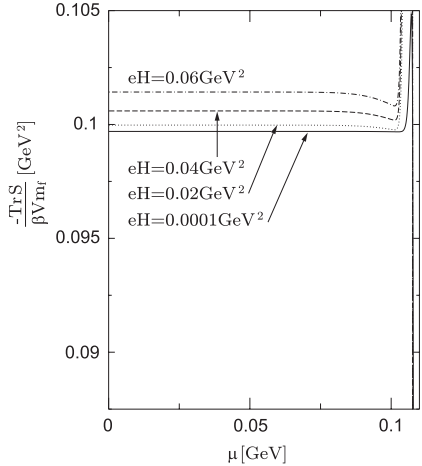}}
 \caption{Behaviors of the vacuum self energy as a function of 
  $\mu$ with $H$ and $T$ fixed. We set $\Lambda=2$GeV,
 $m_s=m_f=0.1$GeV, $T=0.006$GeV and
  $eH=0.0001$GeV$^2$, $eH=0.02$GeV$^2$, $eH=0.04$GeV$^2$, 
  $eH=0.06$GeV$^2$. A vertical line of right side consists of all the 
  value of $eH$.}
 \label{ell}
 \end{center}
\end{figure}

We apply our formalism to the vacuum self-energy at the one-loop 
level. At the limit $T \rightarrow 0$ and/or $\mu \rightarrow 0$
contributions from the poles at $eH\tau = n\pi$ disappear and
our results coincide with the previous one. We numerically 
perform the proper-time integral and the Matsubara mode summation.
In a strong magnetic field oscillating mode is observed for both
the scalar and the fermion fields.
Our results qualitatively agree with the previous analysis 
obtained by the naive Wick rotation$^{13}$ for $\mu < m_f$.
Performing the naive Wick rotation, Eq.(\ref{Tr-Green2}) reads
\begin{equation}
 \frac{1}{\beta V}{\rm TrS} = - \frac{m_f}{2\pi^{3/2}\beta}
  \int^\infty_{1/\Lambda^2} d\tau \frac{eH}{\sqrt{\tau}} 
  \coth(eH\tau) e^{\tau(\mu^2-m_f^2)}\theta_2 
  ( 2\mu\tau / \beta, 4\pi i\tau / \beta^2) \ .
\label{eth}
\end{equation}
In Fig. \ref{ell} we show the behaviors of the fermion vacuum 
self-energy (\ref{eth}).
 $T$ and $\mu$ are fixed on the same value with the 
 Fig. \ref{mu}. We cut the proper-time integration at $10^4$GeV$^{-2}$ 
 because it does not converge for $\mu > m_f$.
 As is shown in Fig. \ref{ell}, the vacuum 
 self-energy coincides with the results in Fig. \ref{mu}
 for $\mu < m_f = 0.1$GeV. But it is completely different 
 for $\mu > 0.1$GeV.
In Fig. 4 the vacuum self-energy is not analytic at some points. 

There are some interesting applications of our work.
Using the two-point functions obtained here we calculate
the radiative correction of some QED process, radiative decay of axion, 
neutrino, and so on.
As an example, we calculated the trace of the two-point
functions. It is necessary to determine the phase structure
of the theory at finite $T$, $\mu$ and $H$.${}^{20}$
The influence of the external magnetic field will modify
the decay rate and phase structure at finite $T$ and $\mu$.
It may affect the evolution of our universe and/or neutron
stars.

The present work is restricted to the analysis of the influence
of the temperature, chemical potential and a constant magnetic
field. There are some interesting objects in an external 
electromagnetic field. However it is not clear to extend our 
procedure to the state in an external electric field.
It is also interesting to introduce the curvature effects
in our analysis.$^{18}$

\section*{Acknowledgements}
The authors would like to thank Takahiro Fujihara and Xinhe Meng 
for useful discussions.

\end{document}